\documentstyle[preprint,aps,eqsecnum]{revtex}
\begin{document}
\preprint{July 1994}
\title
{Order $\hbar$ Corrections to the Classical Dynamics of a Particle with
Intrinsic Spin
Moving in a Constant Magnetic Field}

\author{Patrick L. Nash}
\address
{Division of Earth and Physical Sciences \\
University of Texas at San Antonio \\
San Antonio, TX 78249-0663}
\maketitle

\begin{abstract}
\noindent
$O(\hbar)$ effects that modify the classical orbit of a charged particle
are described for the case of a classical spin-$\frac{1}{2}$ particle
moving in a constant magnetic field,
using a manifestly covariant formalism reported previously.
It is found that the coupling between the momentum and spin
gives rise to a shift in the cyclotron frequency, which is
explicitly calculated.
In addition the orbit is found to exhibit $O(\hbar)$ oscillations along the
axis of the uniform static magnetic field
whenever the gyromagnetic ratio $g$ of the particle is not 2.
This oscillation is found to occur at a frequency $\propto$ $\frac{g}{2} - 1$,
and is an observable source of electric dipole radiation.
\end{abstract}



\section{Introduction}

{\it Closed orbit correction} is a critical component of background
reduction in high energy electron and proton accelerator experimental
detectors.
In a collider ring, tight quality control of the closed orbit is always
essential to the efficient operation of the detectors.

Higher-order non-linear relativistic processes
may measurably effect the
dynamical trajectory of a classical spin-$\frac{1}{2}$ electron
in an electron accelerator or a proton in a
particle beam.
Such effects can in principle
be seen after many orbits in a
long-lived charged particle beam in a storage ring when
conditions for resonance are met.
For example,
a key issue in collider physics is to what extent the interaction of
non-vanishing magnetic field gradients along the
particle's trajectory with the intrinsic magnetic dipole moment of the particle
drives the particle from the ideal design orbit
(the relativistic Stern-Gerlach effect).
This may alter the
depolarizing resonance strengths and widths, and may induce new resonant modes
that effect the spin.
Hence a quantitative description of the contribution of
non-linear relativistic effects
to closed orbit control is desirable.
After over fifty years of study, this subject continues to be an
active research topic \cite{kout:89}~-\cite{courant:80}.
Failure to study these processes
could lead to a serious deficiency in our knowledge
of the details of high-energy classical dynamics. For example,
small innocuous non-linear terms can sometimes lead to
spectacular observable experimental consequences (recall
the hysteresis in cyclotron resonance based on a weak non-linear
relativistic mass effect).

Moreover the recent questions concerning the origin of proton spin
(Ashman, {\it et al} \cite{ashman}, Close and Roberts \cite{close},
Ellis and Karliner \cite{ellis}, Meng Ta-chung, {\it et al} \cite{Ta-c})
and the observation of a strong spin-dependence of high-energy
proton-proton interactions underline the need for a complete
understanding of classical spin.

In this paper we consider the classical motion of a charged particle
with intrinsic spin orbiting in a circle
about a constant magnetic field normal (in the $zero^{th}$ approximation)
to the plane of the orbit. Such fields occur
in an ideal accelerator or are approximated by the stellar magnetic field near
a
 pulsar.
We study the dynamical problem classically to $O(\hbar)$ and derive several new
experimentally
testable predictions of classical dynamics.
We also clarify
a classic result, which states that the ratio of the spin precession
angular frequency to the cyclotron frequency is given by
$1 + \gamma (\frac{g}{2} - 1) \; \; ^{\cite{sakurai}, \cite{ruth:83},
\cite{courant}}$,
where $ g $ is the gyromagnetic ratio of the particle and
$\gamma^{-2} =  1 - \frac{v^{2}}{c^{2}}$.
It is shown that this is a time-averaged result, and derive the time-dependent
expression
in Equation[\ref{truth}].
We shall adopt a classical description throughout this paper so that we may
describe the
electron with the same formalism as the proton. A quantum mechanical
self-consistent description
of an electron moving in the prescribed fields would require including
radiative corrections to the electron-photon vertex in
order to account for the fact that $g \neq 2$ for the electron.
(Not so for a quantum mechanical description of protons, where one does not
hesitate
to introduce a phenomenological gyromagnetic ratio.)
In the classical description, $g$ is a parameter that one sets equal to an
experimentally determined value.
Although a classical description is many times inadequate, in this case we
shall see that the classical description predicts and describes some
effects that are difficult to calculate using relativistic quantum mechanics.

Although not treated in this paper, the method employed to formulate this
dynamical problem can be extended to handle the case of charged-particle beams.
This is important since the actual synchrotron beam is comprised of multiple
bunches of charged particles, with up to $10^{11}$ particles per beam
bunch. The dynamics of the beam is determined by the applied external fields
plus
the electromagnetic fields generated by the particles in the interacting
bunches.
The normalized precessional frequency is shifted by the beam-beam interaction
and these shifts can also be approximated.

\section{Equations of Motion}
\label{sec-dyneqs}

The Lorentz force equations for the four-velocity
$ \frac{dx ^{\alpha}}{d \tau} \stackrel {\rm def}{=} \dot{ x} ^{\alpha} $
coupled with the
Thomas-Frenkel-Bargmann-Michael-Telegdi (BMT) equations for the dynamical
evolution of the Pauli-Lubanski spin vector govern the motion
of a polarized particle
in the limit $\hbar \rightarrow 0$. Here $ \tau $ is the proper time.
These equations have been extended to higher order in $\hbar$.
Although there are many specific models $^{\cite{frenkel}-\cite{bowlin:90}}$,
we shall employ the
dynamical equations derived by Nash $^{\cite{nash:84}}$. This coupled set of
equations
reduces {\it exactly} to the standard BMT and Lorentz force equations in the
limit $\hbar \rightarrow 0$.
Moreover these equations are derived from a Lagrangian using a variational
principle,
which ensures that the dynamical equations do not conflict
with conservation laws.
We turn briefly to a summary of the approach of Ref~\cite{nash:84}.

Units are used in which the speed of light is one. The Minkowski
spacetime metric is $\eta _{\alpha \beta} = diag( 1, 1, 1, -1)$.
The dynamical variables are the particle's position
$ x ^{\alpha}, \; \alpha = 1, \ldots, 4$ in spacetime  and $ \psi $,
which is a real eight-component spinor. $\psi $ may be regarded as a column
vector
comprised of the direct sum of a real four-component Dirac spinor $\lambda$
[transforming
under the {\it real } irreducible representation of $ \overline { SO(3,3) }
$ defined
by Dirac (see Ref~\cite{dirac:63})] and the transpose of another real
four-component
Dirac spinor $ \xi $ that transforms under the inverse irreducible
representation
\begin{equation}
		\psi = \left(	\begin{array}{c}
				\lambda	 \\  \tilde \xi
				\end{array}
			\right),
\end{equation}
where the tilde denotes the transpose.
$\psi$ defines an orthonormal tetrad in Minkowski spacetime
$^{\cite{nash:80}}$.
The members of the tetrad
are constructed from a sum of products of the components of $\psi$.
It is known $^{\cite{nash:90}}$ that $\psi$ may be regarded as a (split)
octonian.

Explicitly, the timelike member of the tetrad is given by
\begin{equation}
        E^{\alpha}_{(4)}(\tau) = - \frac{1}{2}
        \tilde{\psi} \Gamma^{4} \Gamma^{\alpha} \psi,
\label{eq:4vel}
\end{equation}
where the $ \Gamma _{\alpha}$ matrices are real $8 \times 8$ analogs of Dirac's
gamma matrices,
and is parallel to the four-velocity $\dot{x}^{\alpha}$ when
Equation~[\ref{eq:diraceq}]
(below) is satisfied. In passing, we record for later use the definition of the
third member of
the tetrad
\begin{equation}
        E^{\alpha}_{(3)}(\tau) = \frac{1}{2}
        \tilde{\psi} \Gamma^{4}  \Gamma^{7} \Gamma^{\alpha} \psi.
\label{eq:4spin}
\end{equation}
For simplicity of presentation it is assumed that $\psi$ has been
normalized to $\tilde{\psi} \psi = \dot{x}^{4}$. The general case is
obtained by dividing by a normalizing factor proportional to
$\xi \gamma^{5} \lambda$.

The classical set of dynamical variables $\{x^{\alpha}, \psi\}$ are not
all independent. This is because the timelike $E^{\alpha}_{(4)}$
constructed from an arbitrary $\psi$ is in general not
parallel to $\dot{x}^{\alpha}$. But in the free-field case one knows that
only one timelike vector is required for the description of particle
`dynamics.' In order for $\{x^{\alpha}, \psi\}$ to describe a massive particle
with spin,
$\psi$ must be constrained so that $E^{\alpha}_{(4)}$
and $\dot{x}^{\alpha}$ are parallel.
The crucial result we shall use is that the timelike member of the
tetrad can be permanently aligned with the four-velocity $ \dot{x} ^{\alpha} $
of the particle
by imposing the constraint $^{\cite{nash:84}}$
\begin{equation}
		\left( \Gamma _{\alpha} \dot{ x} ^{\alpha}  +
		\sqrt{ - \dot{ x} _{\alpha} \dot{ x} ^{\alpha} }\; \Gamma ^{7}
		\right) \psi = 0.
\label{eq:diraceq}
\end{equation}
This {\it classical} constraint bears a striking resemblance to the quantum
mechanical Dirac equation. This suggests that one may quantize this theory by
``quantizing the constraint'' and postulating minimal coupling.
However we do not treat the problem considered in Section \ref{uniform magnetic
field}
quantum mechanically. To do so one must
consistently take into account higher
order corrections to the magnetic moment of the particle that arise from
radiative corrections to the lepton-photon vertex operator.
In the classical approach one may
use the experimentally measured gyromagnetic ratio in the classical equations
of motion.  One obtains an approximation for the particle's orbit.

The constraint Equation~[\ref{eq:diraceq}] is incorporated into the theory
using a Lagrange multiplier.
If this constraint is satisfied then the third member of
the tetrad may be identified with the Pauli-Lubanski spin vector, and moreover,
the intrinsic electric dipole moment of the particle vanishes in a rest frame
$\Sigma _{\alpha \beta}  \dot{ x} ^{\beta} = 0$,
where $\Sigma ^{\alpha \beta}$ is the spin tensor of the particle defined by
\begin{equation}
	\Sigma ^{\alpha \beta} = - \frac{1}{2} \tilde \psi \Omega \left(
		M^{\alpha \beta}
		\right) \psi.
\label{spin tensor definition}
\end{equation}
Here $\Omega $ is the symplectic form on the spinor manifold and
\begin{equation}
	M^{\alpha \beta} = - \frac{1}{4}[ \Gamma ^{\alpha}, \Gamma ^{\beta}].
\end{equation}

The Lagrangian for the theory is
\begin{equation}
	{\cal L} = -M  \sqrt{ - \dot{ x} _{\alpha} \dot{ x} ^{\alpha} }\;
		- \frac{\hbar}{2} \tilde \psi \Omega \dot {\psi}
		+ e A _{\alpha}  \dot{ x} ^{\alpha}
		+ Lagrange \; multiplier \; term
\end{equation}
where $e$ is the charge of the particle, $A _{\alpha}$ is the electromagnetic
vector potential, and $M$ denotes the effective mass of the particle, which
takes
into account the spin-field interaction energy contribution to the particle's
mass. It is given by
\begin{equation}
		M = m \sqrt{1 - \frac{ g e \hbar}{2 m^{2}}
			\Sigma ^{\alpha \beta} F_{\alpha \beta}}
\end{equation}
where m is the rest mass of the particle
(or simply a parameter with dimension of mass, which
is to be renormalized to the observed rest mass)
and $F_{\alpha \beta}$ is the electromagnetic
field tensor. $F_{\alpha \beta}$ contains a contribution from the radiation
reaction
of the particle. If this contribution to $F_{\alpha \beta}$ is completely
neglected
then the equations of motion admit unphysical solutions in the free-field
limit.

The canonical four-momentum  $p_{\alpha}$ is given by
\begin{equation}
p_{\alpha} = M x_{\alpha} + e A_{\alpha} + \hbar \Sigma_{\alpha \beta}
b^{\beta}
\end{equation}.
the classical version
The equations of motion derived from this Lagrangian are
\begin{equation}
		\hbar \dot{\psi} = -\hbar M^{\alpha \beta} \psi \left(
			\frac{ g e}{4 M}  F_{\alpha \beta}
			+ \dot{x}_{\alpha} b_{\beta} \right)
\label{eq:psidot}
\end{equation}
where
\begin{equation}
		b^{\alpha} = \ddot{x}^{\alpha}
		- \frac{g e }{2 M} F^{\alpha}_{ \;\; \beta} \dot{x}^{\beta}.
\end{equation}
Also
\begin{equation}
	M \ddot{x}^{\alpha} = e F^{\alpha}_{\; \; \beta} \dot{x}^{\beta}
		+ P ^{\alpha \mu} \Upsilon _{\mu},
\label{eq:lorentz}
\end{equation}
where $P ^{\alpha \mu}$ is a projection operator
$P ^{\alpha \mu} = \eta ^{\alpha \mu} + \dot{x}^{\alpha}  \dot{x}^{\mu}$
and
$
	\Upsilon _{\mu} = \frac{g e \hbar}{4 M} \Sigma ^{\beta \lambda}
	\frac {\partial F_{\beta \lambda}} {\partial x ^{\mu}}
	- \frac {d}{d \tau} \left( \hbar \Sigma _{\mu \beta} b ^{\beta} \right)
$
contains the contribution to the force due to the Stern-Gerlach effect.

For completeness we mention that upon substituting the result
Equation~[\ref{eq:psidot}]
into the definition of the tetrad $E^{\alpha} _{(j)}$ yields
\begin{equation}
		\dot{E}^{\alpha}_{(j)} =
		\frac{g e}{2 M} F^{\alpha}_{\; \; \beta}E^{\beta}_{(j)}
		+ \dot{x}^{\alpha} b _{\beta} E^{\beta}_{(j)}.
\label{eq:tetraddot}
\end{equation}
where $\; j = 1, 2, 3$ labels the spacelike members of the tetrad.
It is worth noting that in the limit $\hbar \rightarrow 0$
Equation~[\ref{eq:lorentz}]
is the Lorentz force equation and Equation~[\ref{eq:tetraddot}], with $j = 3$,
is
exactly the BMT equation.

\section{Uniform Static Vertical Magnetic Field}
\label{uniform magnetic field}

In order to simplify the notation we define two sets of three $8 \times 8$
matrices
\begin{equation}
                {\bf \sigma} = (M^{23}, M^{31}, M^{12})
\end{equation}
and
\begin{equation}
                {\bf m} = (M^{14}, M^{24}, M^{34}).
\end{equation}
We employ the representation of the
$\Gamma$ matrices defined in References \cite{nash:90} and \cite{nash:86}.

The electromagnetic field tensor $F_{\alpha \beta}$ is represented
in terms of the electric field ${\bf E}$ and
magnetic field ${\bf B}$ as
\begin{equation}
F_{\alpha \beta}  = \left( \begin{array}{cccc}
		0 & B_{3} & - B_{2} & E_{1} \\
		- B_{3} & 0 & B_{1} & E_{2} \\
		B_{2} & - B_{1} & 0 & E_{3} \\
		- E_{1} & - E_{2} & - E_{3} & 0
		\end{array} \right)
\end{equation}

We assume that the $\hbar \rightarrow 0$ limit of the Lorentz force equations
have been solved so
that $x^{\alpha} = x^{\alpha}(\tau)$ is known. Substituting this result
into the $\left. \frac{d}{d \hbar} \right|_{\hbar \rightarrow 0}$ limit of
Equation~[\ref{eq:psidot}] yields
\begin{equation}
	\dot{\psi} = V(\tau)\psi,
\label{eq:psi}
\end{equation}
where
\begin{equation}
	V = \frac{e}{2 m}[ - \frac{g}{2}F_{\alpha \beta} +
        (\frac{g}{2} - 1)\dot{x}^{\mu}(\dot{x}_{\alpha}F_{\beta \mu} -
        \dot{x}_{\beta}F_{\alpha \mu})] M^{\alpha \beta}
\label{eq:vee}
\end{equation}
is real $8 \times 8$ matrix that is a known function of tau.

In the notation introduced above Equation~[\ref{eq:vee}] reduces to
\begin{eqnarray}
\frac{m}{e}V & = & -\{ 1 + \gamma^{2} (\frac{g}{2} - 1) \} {\bf \sigma \cdot B}
+ (\frac{g}{2} - 1) \{ {\bf \dot{x} \cdot B} \} \;
{\bf \sigma \cdot \dot{x}} \nonumber \\
     & & \mbox{} - \gamma (\frac{g}{2} - 1) {\bf m \cdot B \times \dot{x}}
      - (\frac{g}{2} - 1)\{ {\bf \dot{x} \cdot E} \} \; {\bf m \cdot \dot{x}}
\nonumber \\
     & & \mbox{} - \gamma^{2} (1 - \beta^{2} \frac{g}{2}){\bf m \cdot E}
      + \gamma (\frac{g}{2} - 1) {\bf \sigma \cdot \dot{x} \times E}.
\label{eq:v}
\end{eqnarray}

For the limiting case
$\hbar \rightarrow 0, \; {\bf B} = {\bf \hat{z}}B_{3}, \; B_{3} = constant$,
the electromagnetic field has only one non-vanishing
independent component in the lab frame. This is $F_{12} = B_{3}$.
The Lorentz force equations~[\ref{eq:lorentz}] are independent of $\psi$ to
$zero^{th}$ order in $\hbar$.
The $zero^{th}$ order solution $\hbar \rightarrow 0$ is of course well known.
For a
circular orbit in the ${\bf x} \wedge {\bf y}$ plane a solution to the Lorentz
force equation is $\dot{x}^{1} = \beta \gamma \cos (\omega_{0} \tau)$
, $\dot{x}^{2} = - \beta \gamma \sin(\omega_{0} \tau)$,
$\dot{x}^{3} = 0$ and
$\dot{x}^{4} = \gamma = constant$,
where $\omega_{0} = \frac{e B_{3}} {m} \;$,
$\beta = \frac{v}{c} = v = constant$
and $\gamma^{2} = \frac{1}{1 - \beta^{2}}$.
The cyclotron frequency is
$\omega_{{\it cyclotron}} = \mid \omega_{0} \mid / \gamma$.

If we put $T = \omega_{0} \tau$ and $\nu = \gamma (\frac{g}{2} - 1)$,
and use Equation~[\ref{eq:v}], then Equation~[\ref{eq:psi}] yields
\begin{equation}
\frac{d \psi}{dT} = - \{ (1 + \gamma \nu) M^{12} +
\beta  \gamma \nu [\sin(T) M^{14} + \cos (T) M^{24}]\} \psi.
\label{eq:v0}
\end{equation}
Upon making the substitution
\begin{equation}
        \phi = \exp [ (1 + \gamma \nu) T M^{12}] \psi,
\label{eq:phi}
\end{equation}
and simplifying, one finds that $\phi$ satisfies
\begin{eqnarray}
\frac{d \phi}{dT} & = & -\beta \gamma \nu  [- \sin(\gamma \nu T) M^{14}
+ \cos(\gamma \nu T) M^{24}] \phi \nonumber \\
 & = & -\beta \gamma \nu e^{\gamma \nu
TM^{12}}M^{24}e^{- \gamma \nu TM^{12}} \phi,
\label{eq:phidot}
\end{eqnarray}
where we have used
\begin{eqnarray}
e^{(1 + \gamma \nu) T M^{12}} [ \sin(T) M^{14} + \cos (T) M^{24} ]
e^{-(1 + \gamma \nu)T M^{12}} & & \nonumber \\
= - \sin(\gamma \nu T) M^{14} + \cos (\gamma \nu T) M^{24}.
\end{eqnarray}

The general solution to Equation~[\ref{eq:phidot}] is now easily found.
One finds that $\psi = \Lambda_{{\bf B_{3}},g,\beta,\tau}\; \psi(0)$, where
the constants of integration are in $\psi(0)$ and
\begin{eqnarray}
\Lambda_{{\bf B_{3}},g,\beta,\tau} & = & \
               e^{-T M^{12}} e^{- \gamma \nu T (M^{12} + \beta M^{24})}
\nonumber \\
         & = & e^{-\kappa [ \cos (T) M^{14} -
\sin(T) M^{24}]} e^{- (1 + \nu)T M^{12}} e^{\kappa M^{14}},
\label{eq:Szero}
\end{eqnarray}
and
$\kappa =  \frac{1}{2}\ln \{ \frac{1 + \beta}{1 - \beta} \}$. $e^{\kappa}$ is
Bondi's `k' factor $^{\cite{bondi:64}}$.
$\Lambda_{{\bf B_{3}},g,\beta,\tau}$ is decomposed into the product of a boost
in the direction of $-{\bf \dot{x}}(0)$,
times a $\tau$-dependent rotation about the guiding magnetic field,
times another boost in the
${\bf \hat{x}}\cos(T) - {\bf \hat{y}} \sin(T) \propto {\bf \dot{x}}(T)$
direction.
$\Lambda_{{\bf B_{3}},g,\beta,\tau}$
commutes with $\Gamma^{3}$ and hence does not affect the polarization
$\Sigma_{3}$.

In order to calculate the Pauli-Lubanski spin vector from
$\psi(\tau) = S_{0} \psi(0)$, with $S_{0} = \Lambda_{{\bf B_{3}},g,\beta,\tau}$
we first solve the eigenvalue problem Equation~[\ref{eq:diraceq}] for
$\psi(0)$,
choosing initial phases that simplify our work:
\begin{equation}
                \left( \Gamma _{\alpha} \dot{ x} ^{\alpha}(0)  +
                        \sqrt{ - \dot{ x} _{\alpha}(0) \dot{ x} ^{\alpha}(0)
}\; \Gamma ^{7}
                \right) \psi(0) = 0.
\end{equation}
We find that there are four linearly independent solutions to this eigenvalue
problem for $\dot{x}^{4} > 0$, two solutions for spin up, and two solutions
with spin down.
As a check on our work we
substitute $\psi(0)$ and $S_{0}$ into the definition
$E^{\alpha}_{(4)}(\tau) = - \frac{1}{2}
        \tilde{\psi}(0)\tilde{S}_{0} \Gamma^{4} \Gamma^{\alpha} S_{0} \psi(0)$.
Each distinct spinor solution,
when substituted into this equation yields the four-velocity
$E^{\alpha}_{(4)} = (\beta \gamma cos(T), -\beta \gamma sin(T), 0,
\gamma) = \dot{x}^{\alpha}$,
as required. Next,
substitution of the four spinor eigenvectors in turn into the definition
Equation~[\ref{eq:4spin}]
yields two distinct Pauli-Lubanski four-vectors, which differ only in the sign
of
the third component. For example, we find that a ``spin up'' spinor eigenvector
is
\begin{equation}
\tilde{\psi}(0)_{\uparrow} = (0,0,1,0,0,-\beta \gamma,0,\gamma)/\sqrt{2
\gamma},
\label{eq:third spinor eigenvector}
\end{equation}
and
$\psi = \Lambda_{{\bf B_{3}},g,\beta,\tau}\; \psi(0)_{\uparrow} = $ is given by
\begin{equation}
	\psi = \frac{1}{\sqrt{2 \gamma}} \left(
			\begin{array}{c}
	       		- \beta \gamma \sin(T/2) \sin(\nu T/2) \\
	       		- \beta \gamma \cos(T/2) \sin(\nu T/2) \\
	       \cos(T/2) \cos(\nu T/2) - \gamma \sin(T/2) \sin(\nu T/2) \\
              - \sin(T/2) \cos(\nu T/2) - \gamma \cos(T/2) \sin(\nu T/2) \\
                        - \beta \gamma \sin(T/2) \cos(\nu T/2) \\
                             - \beta \gamma \cos(T/2) \cos(\nu T/2) \\
	       		\cos(T/2) \sin(\nu T/2) + \gamma \sin(T/2) \cos(\nu T/2) \\
                      - \sin(T/2) \sin(\nu T/2) + \gamma \cos(T/2) \cos(\nu
T/2)
			\end{array} \right)
\label{eq:time dep spinor eigenvector}
\end{equation}
The Pauli-Lubanski spin four-vector in this case is
\begin{eqnarray}
	E^{\alpha}_{(3)} & = &
( -\frac{1}{2} \beta [ (\gamma + 1) \cos(\nu + 1)T + (\gamma - 1)
\cos(\nu - 1)T], \nonumber \\
	& & \frac{1}{2} \beta [ (\gamma + 1) \sin(\nu + 1)T - (\gamma - 1)
\sin(\nu - 1)T], \nonumber \\
	& & \gamma^{-1}, -\beta^{2} \gamma cos\nu T).
\label{eq:spin_vector}
\end{eqnarray}
$E^{\alpha}_{(3)}$ is just the Lorentz transform to the lab frame
of the spin 3-vector
$
{\bf s} = ( - \beta \cos[(\nu + 1) T] , \beta \sin[(\nu + 1) T] ,
\frac{1}{\gamma} )$,
which may be verified by transforming $E^{\alpha}_{(3)}$
to an instantaneous rest frame with the Lorentz boost $L_{boost}$ defined as
usual by
$
x'^{i} = x'^{j} ( \delta^{i}_{j} - n^{i}n_{j} + n^{i}n_{j} ) =
( \delta^{i}_{j} - n^{i}n_{j} ) x^{j} + n^{i} \gamma ({\bf \hat{n}} \cdot
\vec{x} - \beta c t)
\equiv L_{boost } \, ^{i}_{\; \alpha} x^{\alpha}
$
and
$
c t' = \gamma ( c t - \beta {\bf \hat{n}} \cdot \vec{x} )
\equiv L_{boost } \, ^{4}_{\; \alpha} x^{\alpha},
$
where $n^{i}$ is a unit vector parallel to the 3-velocity of the particle.
Applying this
boost to $E^{\alpha}_{(3)}$ yields
\begin{equation}
	E^{rest \; frame} \; ^{\alpha}_{(3)} = ( - \beta \cos[(\nu + 1) T] ,
\beta \sin[(\nu + 1] T) ,
			\frac{1}{\gamma} ,0 ).
\end{equation}
Since $\dot{x}_{\alpha} E^{\alpha}_{(3)} = 0$,
the angle $\theta_{R}$ between the three-velocity
$\dot{x}^{j}$ and the three-spin $E^{j}_{(3)}$ is $\nu T$.
One sees that $\theta_{R} \stackrel{def}{=}
\nu T = \nu \omega_{0} \tau = \gamma (\frac{g}{2} - 1) \frac{e B_{3}}{m}
\frac{t}{\gamma}$,
which is the well known result for the precession of the
longitudinal polarization $^{\cite{hage:73}}$.

The angular velocity of precession $\omega_{{\it precess}}$
that is measured in the lab frame is given by
\begin{eqnarray}
	\omega_{{\it precess}}/\omega_{{\it cyclotron}} & = & \left|
(E_{(3)}^{1}\frac{d E_{(3)}^{2}}{d T}
		- E_{(3)}^{2}\frac{d E_{(3)}^{1}}{d T})/
		(E_{(3)}^{1})^{2} + (E_{(3)}^{2})^{2}
		\right| \nonumber \\
	& = & 1 + \frac{g/2 - 1}{1 - \beta^{2} \sin^{2}(\nu T)}.
\label{truth}
\end{eqnarray}
The average over a time $\nu T = 2 \pi$ is
\begin{equation}
	< \omega_{{\it precess}}/\omega_{{\it cyclotron}} > = 1 +
\gamma (\frac{g}{2} - 1) = 1 + \nu.
\end{equation}
which is the well-known expression often identified as the normalized spin
precessional
frequency (see, for example, References \cite{sakurai}, \cite{ruth:83} and
\cite{courant}).

In order to write out and solve the momentum equations [\ref{eq:lorentz}] to
$O(\hbar)$
we must first evaluate the components of the spin tensor. Substituting $\psi$
form Equation~[~\ref{eq:time dep spinor eigenvector}~] into
Equation~[~\ref{spin tensor definition}~] yields
\begin{equation}
\left( \begin{array}{c} \Sigma^{23} \\ \Sigma^{31} \\ \Sigma^{12} \\
			\Sigma^{14} \\ \Sigma^{24} \\ \Sigma^{34}
\end{array} \right)
=
\left( \begin{array}{c}
		\beta [- \cos(T) \cos(\nu T) + \gamma \sin(T) \sin(\nu T)] \\
		\beta [\sin(T) \cos(\nu T) + \gamma \cos(T) \sin(\nu T)] \\
		1 \\
		- \beta \sin(T) \\
		- \beta \cos(T) \\
		- \beta^{2} \gamma \sin(\nu T)
\end{array} \right) .
\end{equation}
In terms of their Fourier decompositions,
$\Sigma^{23} = \frac{\beta}{2} [ - (\gamma + 1) \cos((\nu + 1) T) - (\gamma -
1) \cos((\nu - 1) T)$ and
$\Sigma^{31} = \frac{\beta}{2} [   (\gamma + 1) \sin((\nu + 1) T) + (\gamma -
1) \sin((\nu - 1) T)$.

A word about normalization. Throughout this paper
the normalization of $\psi$ is determined by the arbitrary requirement
that the $E^{\alpha}_{(\beta)}$ comprise a {\it normalized} set of four
mutually orthogonal vectors,
which is reflected in the result that $\Sigma^{12} = 1$. This is not a
statement about the
magnitude of the classical spin.
The magnitude of the classical spin is not predicted by this theory.
Instead it must be imposed as an initial condition; in virtue of the equations
of motion
this magnitude is a constant of the motion.
Given this magnitude and $g$ we may employ this formalism to compute
the trajectory. Conversely, one can impose the observed spin magnitude as an
initial
condition and then use the predictions of this theory and
experimental measurements to determine $g$.
One sees that to correctly apply this dynamical formalism to
a spin-$s$ particle one must arrange that $\Sigma^{12} = s$.
This is easy to accomplish by imposing this as an initial condition,
and is manifested in the simple replacement
$\psi(0) \mapsto \sqrt{s} \psi(0)$, which we shall henceforth employ.

We turn now to the $O(\hbar)$ solution of the extended Lorentz
equations for the momentum.
The effective mass
$M = m \sqrt{1 - \frac{g e \hbar}{2 m^{2}} \Sigma^{\alpha \beta}
F_{\alpha \beta}}
\approx m - s \frac{g}{2} \hbar \omega_{0}$ to $O(\hbar)$.
Also to this order,
$\hbar b^{\alpha} = - G \hbar \omega_{0} F^{\alpha}_{\; \; \beta}$
$_{0}\dot{x}^{\beta}/B_{3}$, where $G = \frac{g}{2} - 1$
and $_{0}\dot{x}^{\beta}$ refers to the $O(1) \stackrel{\rm def}{=}
\hbar \rightarrow 0$ solution
to the momentum equations. We find that
\begin{equation}
\hbar \Sigma^{\alpha}_{\; \;  \beta} b^{\beta} = s
\beta \gamma \hbar \omega_{0} G \left( \begin{array}{c}
	\cos(T) \\ - \sin(T) \\ \beta \cos(\nu T) \\ \beta
\end{array} \right)
= s \hbar \omega_{0} G \left( \begin{array}{c}
_{0}\dot{x}^{1} \\  _{0}\dot{x}^{2} \\ \beta^{2}
\gamma \cos(\nu T) \\ \beta^{2} _{\; \;\; 0}\dot{x}^{4}
\end{array} \right) .
\end{equation}
We note that $_{0}\dot{x}^{\alpha} \Sigma_{\alpha \beta} b^{\beta} = 0 = $
$_{0}\dot{x}^{\alpha} \frac{d}{d \tau} \left( \Sigma_{\alpha \beta} b^{\beta}
\right)$.

We write $\dot{x}^{\alpha} = $ $_{0}\dot{x}^{\alpha} +
\hbar$ $_{1}\dot{x}^{\alpha}$ and solve
$\frac{d}{d \tau} \left( M \dot{x}^{\alpha} \right) =
e F^{\alpha}_{\; \; \beta} \dot{x}^{\beta} -
\hbar P^{\alpha \mu} \frac{d}{d \tau} \left( \Sigma_{\mu \beta}
b^{\beta} \right)$.
Substituting for $P^{\alpha \mu}$ and rearranging terms yields
$\frac{d}{d \tau} \left( M \dot{x}^{\alpha} + \hbar \Sigma^{\alpha}_{\; \;
\beta} b^{\beta} \right) =
e F^{\alpha}_{\; \; \beta} \dot{x}^{\beta} -
\hbar \dot{x}^{\alpha} \dot{x}^{\mu} \frac{d}{d \tau} \left( \Sigma_{\mu \beta}
b^{\beta} \right) =
e F^{\alpha}_{\; \; \beta} \dot{x}^{\beta} -
\hbar$ $_{0}\dot{x}^{\alpha}$ $_{0}\dot{x}^{\mu} \frac{d}{d \tau}
\left( \Sigma_{\mu \beta} b^{\beta} \right) =
e F^{\alpha}_{\; \; \beta} \dot{x}^{\beta}$
since
$_{0}\dot{x}^{\alpha} \frac{d}{d \tau} \left( \Sigma_{\alpha \beta} b^{\beta}
\right) = 0$.
For $\alpha = i = 1,2$ we see that
\begin{equation}
\left( m - s \hbar \omega_{0} \right) \ddot{x}^{i} = e F^{i}_{\; \; \beta}
\dot{x}^{\beta}.
\end{equation}
This is independent of $g$, but dependent on the magnitude of the spin,
and leads to a shift in the cyclotron frequency.
This shift is in the other direction for a particle with spin down.
For a particle with $s = \frac{1}{2}$ one sees that
$\gamma \omega_{{\it cyclotron}} =  \left| \frac{ e B_{3}}{m - \frac{1}{2}
\hbar \omega_{0}} \right| =
\frac{\omega_{0}}{1 - \frac{1}{2} \frac{ \hbar \omega_{0}}{m}}$. This
effect may be of interest for the case of plasma dynamics in
critical magnetic fields of the order of $10^{12} - 10^{14} Gauss$, which may
exist near pulsars.

Continuing with the analysis we find that $\dot{x}^{4} = \gamma = constant$ and
\begin{eqnarray}
\dot{x}^{3} & = & - \beta^{2} \frac{\hbar s \omega_{0}}{m} \nu \cos(\nu T)
+ constant \nonumber \\
	& = & - \frac{d}{d \tau} \left( \beta^{2} \frac{\hbar s}{m}
\sin(\nu T) \right),
\end{eqnarray}
where for simplicity we have set the integration constant equal to zero.
The charge will radiate in virtue of this oscillation along the
axis of the applied magnetic field. The oscillation exists because
the four-spin is orthogonal to the four-velocity, and the spin precesses.

\section{Conclusion}
\label{conclusion}

The theory predicts a contribution to the electric dipole radiation of the
particle
with frequency
$\omega = \omega_{0} ( \frac{g}{2} - 1)$ and power on the order of
$
\frac{1}{3} c k^{4} \left| \vec{p} \right| ^{2}
=
\frac{\omega^{4} }{6 c^{3} } \left( \beta^{2} e \frac{\hbar}{2 m c} \right)^{2}
$
due to the $O(\hbar)$ oscillations of the orbit along the direction of the
applied
static magnetic field.
(There are of course other $O(\hbar)$ sources of electric dipole radiation that
we are not
considering here. One such source is proportional to
$ \hbar \partial_{\alpha} \frac{\Sigma^{\alpha \beta}}{M}$.)
This suggests a new way to measure the gyromagnetic ratio of the electron or
proton.
One can measure $\frac{g}{2} - 1$ by measuring the frequency of
the electric dipole radiation due to the $O(\hbar)$ oscillations of the orbit
along the direction of the applied static magnetic field.

Radiation reaction terms have not been included in the analysis.
However, radiation damping tends to polarize electrons in high energy
accelerators.
These effects are observable and must be included in any complete
description of relativistic dynamics of light particles.
The radiation damping force is usually assumed to
arise from the self-field of the accelerated charged particle
{\it via} the Lorentz force.
Dirac has shown that the finite (on the world line of the particle)
contribution to the self-field is given by
$F^{\alpha \beta}_{rad} =
\frac{1}{2} \left( F^{\alpha \beta}_{ret} - F^{\alpha \beta}_{adv} \right) =
\frac{2 e}{3} \left(
v^{\alpha} \ddot{v}^{\beta} - v^{\beta} \ddot{v}^{\alpha} \right) =
\frac{2 e}{3}
\frac{d}{d \tau} \left( v^{\alpha} \dot{v}^{\beta} - v^{\beta} \dot{v}^{\alpha}
\right)$,
where $v^{\alpha} = \dot{x}^{\alpha}$. Adding this field to the externally
applied
electromagnetic field $F^{\alpha \beta}$ in Equation~[~\ref{eq:lorentz}~]
yields a generalization
of the Abraham-Lorentz equation. Such equations are well known to be plagued
with
``self accelerated'' runaway solutions. One simple way around this is to
replace
$\dot{v}^{\alpha}$ with $\frac{e}{M} F^{\alpha}_{\; \; \beta} v^{\beta}$, since
an external $F^{\alpha \beta}$ is ultimately responsible for
$\dot{v}^{\alpha} \neq 0$.
This yields a second order ODE for $x^{\alpha}$ (coupled to the first order
ODE for $\psi$).



\end{document}